\documentclass[aps,prl,groupedaddress]{revtex4}
\usepackage{amsmath}
\usepackage{epsfig,epsf}
\usepackage{graphicx}
\usepackage{verbatim}
\def\eq#1{{Eq.~(\ref{#1})}}
\def\fig#1{{Fig.~\ref{#1}}}
\begin{document}

%\hoffset-1cm

% Yale printer values
\voffset1.5cm
\title{Inside looking out: probing JIMWLK with BFKL calculations.}
\author{Tolga Altinoluk$^1$, Alex Kovner$^1$ and Eugene Levin$^{2,3}$}
\affiliation{$^1$Physics Department, University of Connecticut, 2152 Hillside road, Storrs, CT 06269, USA\\
$^2$ Departamento de F\'\i sica, Universidad T\'ecnica
Federico Santa Mar\'\i a, Avda. Espa\~na 1680,
Casilla 110-V,  Valparaiso, Chile \\
$^3$ Department of Particle Physics,  Tel Aviv University , Tel Aviv 69978, Israel}
 
\date{\today}
\begin{abstract}
We investigate the relation between the eigenvalues and eigenfunctions of the BFKL and JIMWLK/KLWMIJ Hamiltonians. We show that the eigenvalues of the BFKL Hamiltonians are also {\it exact} eigenvalues of the KLWMIJ (and JIMWLK) Hamiltonian, albeit corresponding to possibly non normalizable eigenfunctions. The question whether a given eigenfunction of BFKL corresponds to a normalizable eigenfunction of KLWMIJ is rather complicated, except in some obvious cases, and requires independent investigation. As an example to illustrate this relation we concentrate on the color octet exchange in the framework of KLWMIJ Hamiltonian. We show that it corresponds to the reggeized gluon exchange of BFKL, and find first correction to the BFKL wave function, which has the meaning of the impact factor for shadowing correction to the reggeized gluon. We also show that the bootstrap condition in the KLWMIJ framework is satisfied automatically and does not carry any additional information to that contained in the second quantized structure of the KLWMIJ Hamiltonian. This is an example of how the bootstrap condition inherent in the $t$-channel unitarity, arises in the $s$-channel picture.
\end{abstract}
\maketitle

%%%%%%%%%%%%%%%%%%%%%%%%%%%%%%%%%%%%%%%%%%%%%%%%%%%%%%%%%%%%%%%%%%%%%%
\section{Introduction}
In this paper we study the relation between the BFKL perturbative resummation of leading logarithms \cite{bfkl} at high energy and the JIMWLK/KLWMIJ evolution \cite{jimwlk},\cite{klwmij} equations, which take into account the physics of saturation and multiple scatterings \cite{GLR}. With a slight abuse of language we call BFKL the second quantized formulation which leads not only to the BFKL equation for two gluon exchange, but also to the whole set of BKP equations for $t$-channel exchanges with an arbitrary number of gluons \cite{bkp}.
The basic correspondence between the elements of the two approaches was established in \cite{reggeon} and we will review it briefly below.

 Here we are interested in the question how to relate the eigenvalues and the eigenfunctions of the BFKL Hamiltonian and their counterparts in the JIMWLK/KLWMIJ theory. We will show that the eigenfunctions of the JIMWLK/KLWMIJ Hamiltonians when expanded in Taylor series in the appropriate variable ($\rho$ for JIMWLK and $\delta/\delta\rho$ for KLWMIJ) are eigenfunctions of the BFKL Hamiltonian. However most of the BFKL eigenfunctions when resummed into solutions of JIMWLK/KLWMIJ are not normalizable. The question which are, and which are not, cannot be settled within the BFKL framework. We also discuss how to calculate higher corrections to the BFKL eigenfunctions. As an example we concentrate on the reggeized gluon \cite{reggeization}. We calculate corrections to the reggeized gluon wave function. We show that terms with up to two gluons in the $t$-channel reggeize due to the bootstrap condition inherent in this approach. We also point out that in the JIMWLK/KLWMIJ framework the bootstrap is a necessary consequence of the hermiticity of the JIMWLK/KLWMIJ Hamiltonian, and does not require a specific form of the emission kernel. We calculate corrections to the eigenfunction beyond the two gluon exchange approximation and thereby find the screening correction which appears when at least three gluons are exchanged in the $t$-channel. This is an interesting example of the interrelation between the $t$-channel unitarity, which is the origin of Reggeons and the $s$-channel one which is the inherent feature of the JIMWLK/KLWMIJ approach.
 
\begin{figure}
\includegraphics{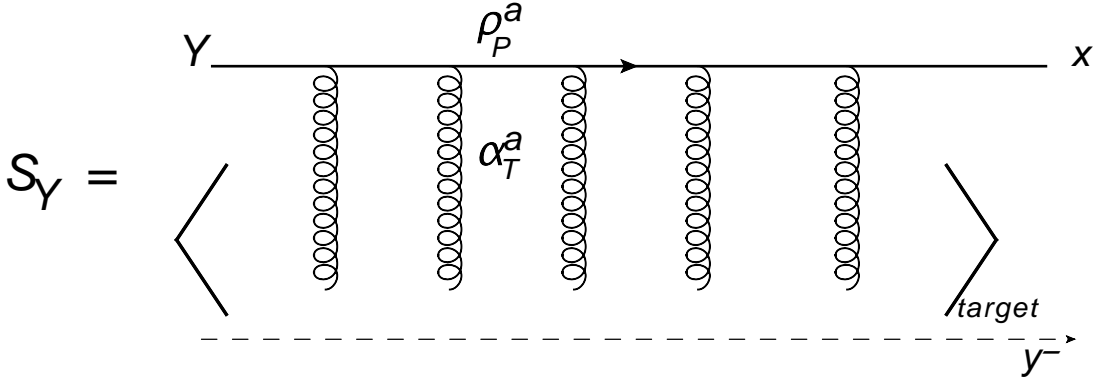}
\caption{\label{btstrp1}The $S$ matrix in the JIMWKL approach with averaging specified by \eq{smatrix}.}
\end{figure}

Let us start by recapitulating the JIMWLK/KLWMIJ formalism. In this approach 
one considers the scattering of a projectile hadron on a hadronic target. The projectile is described by a distribution of color charge density in the transverse plane $\rho_P(x)$, while the target is viewed as an ensemble of the color fields $\alpha_T(x)$ with probability densities $W^P[\rho_P]$ and $W_T[\alpha_T]$ respectively.
The second quantized $S$ matrix operator is given by its eikonal expression (see \fig{btstrp1})
\begin{equation}
\hat S=\exp\left\{i\int_{0}^{1} dy^-\int d^2x\,\hat\rho_P^a(x,y^-)\,\hat\alpha_T^a(x,y^-)\right\}
\end{equation}
and the forward $S$ matrix element at rapidity $Y$ is given by the functional integral
\begin{equation}\label{smatrix}
S_Y=\,\,\int\, D\alpha_T^{a}\,\, \int d\rho_P\,\,W_Y^P[\rho_P]\,\,W^T[\alpha_T]
\,\exp\left\{i\int_{0}^{1} dy^-\int d^2x\,\rho_P^a(x,y^-)\,\alpha_T^a(x,y^-)\right\}
\end{equation}
%%%%%%%%%%%%%%%%%%%%%%%%%%%%%%%%%%%%%%%%%%%%%%%%%%%%%%%%%%%%%%%%%%%%
%%%%%%%%%%%%%%%%%%%%%%%%%%%%%%%%%%%%%%%%%%%%%%%%%%%%%%%%%%%%%%%%%%%%

The rapidity evolution of basic physical observables (including the $S$-matrix) is determined by the second quantized evolution Hamiltonian $H_{RFT}$ (the following pertains to the projectile evolution - we drop the subscript on $\rho$ for brevity)
\begin{equation}\label{evoleq}
-\frac{d}{dY}W[\rho]=H_{RFT}[\rho, \frac{\delta}{\delta\rho}]W[\rho]
\end{equation}
The evolution Hamiltonian $H_{RFT}$  depends on two unitary matrices, 
\begin{equation}
S(x)\,\,=\,\,{\cal P}\,\exp\left\{i\int_{0}^{1} dy^-\,T^a\,\alpha^a(x,y^-)\right\} \ \ \ \ \ 
R(x)\,=\,{\cal P}\exp\left\{\int_0^1 dx^-\, \frac{\delta}{\delta\rho^a(x,x^-)}\,T^a\right\}\,.
\label{alpha1}
\end{equation}
Here the field $\alpha$ is the projectile color field analogous to $\alpha_T$. It is related by a nonlinear transformation to the projectile color charge density $\rho$
\begin{equation}\label{alpha}
\alpha^a(x,x^-)T^a\,\,=g^2\,\,\frac{1}{\partial^2}(x-y)\,
\left\{S^\dagger(y,x^-)\,\,\rho^{a}(y,x^-)\,T^a\,\,S(y,x^-)\right\}\,.
\end{equation}
with $\frac{1}{\partial^2}(x,y)\,=\,\frac{1}{2\pi}\, \ln [(x-y)^2\mu^2]$ and 
\begin{equation}
S(x, x^-)\,\,=\,\,{\cal P}\,\exp\left\{i\int_{0}^{x^-} dy^-\,T^a\,\alpha^a(x,y^-)\right\}
\end{equation}

The $S$-matrix at arbitrary rapidity can therefore be represented in terms of the eigenvalues of $H_{RFT}$. 
Its evolution is given by
\begin{equation}\label{smatrix1}
\frac{dS_Y}{dY}=-\,\,\int\, D\alpha_T^{a}\,\, \int d\rho_P\,\,H_{RFT}[\rho,\frac{\delta}{\delta\rho}]W_Y^P[\rho]\,\,W^T[\alpha_T]
\,\exp\left\{i\int_{0}^{1} dy^-\int d^2x\,\rho^a(x,y^-)\,\alpha_T^a(x,y^-)\right\}
\end{equation}

Given the set of complete eigenfunctionals of $H_{RFT}$
\begin{equation}
H_{RFT}\Psi_i[\rho]=\omega_i\Psi_i[\rho]
\end{equation}
one can expand the probability distribution at the initial rapidity $Y_0$ as
\begin{equation}\label{psii}
W_{Y_0}[\rho]=\sum_i\gamma_i\Psi_i[\rho]
\end{equation}
and therefore
\begin{equation}\label{psiii}
S_Y=\sum_i e^{-\omega_i(Y-Y_0)}\gamma_i\beta_i
\end{equation}
with
\begin{equation}
\beta_i=\int\, D\alpha_T^{a}\,\, \int d\rho\,\,\Psi_i[\rho]\,\,W^T[\alpha_T]
\,\exp\left\{i\int_{0}^{1} dy^-\int d^2x\,\rho^a(x,y^-)\,\alpha_T^a(x,y^-)\right\}
\end{equation}

Given that the Hamiltonian $H_{RFT}$ is positive definite \cite{yinyang}, the asymptotic behavior of the $S$-matrix is governed by the lowest lying eigenvalues $\omega_i$. The study of the spectrum of $H_{RFT}$ is therefore of direct relevance to understanding the high energy behavior of physical amplitudes. 

Two limits of Hamiltonian $H_{RFT}$ have been widely discussed in the literature. One 
is valid in the limit of dense projectile, but allows only exchange of two gluons with the target \cite{jimwlk}. In the other limit one assumes that the projectile is dilute, but resumms all possible multiple interactions of the partons of the projectile with the target \cite{klwmij}. 
The two limiting cases are related by the dense-dilute duality transformation \cite{duality} and have therefore identical spectra. In the rest of this paper we choose to study the KLWMIJ Hamiltonian.
\begin{equation}
H_{KLWMIJ}=\frac{\alpha_s}{2\pi^2}\int_{x,y,z}{K_{xyz}\left\{J^a_L(x)J^a_L(y)+J^a_R(x)J^a_R(y)-2J^a_L(x)R^{ab}_zJ^b_R(y)\right\}}
\label{klwmij}
\end{equation}
with the kernel
\begin{equation}
K_{xyz}=\frac{(x-z)_i(y-z)_i}{(x-z)^2(y-z)^2}
\end{equation}
and 
the left and right rotation generators
\begin{eqnarray}\label{LR}
J^a_L(x)=-tr\left[\frac{\delta}{\delta R^{\dagger}_x}T^aR_x\right]  \\  
J^a_R(x)=-tr\left[R_xT^a\frac{\delta}{\delta R^{\dagger}_x}\right] 
\end{eqnarray}

Alternatively one can write
\begin{equation}
H_{KLWMIJ}=\frac{\alpha_s}{2\pi^2}\int_{x,y,z}{K_{xyz}\left\{J^a_V(x)J^a_V(y)-2J^a_L(x)\left(R_z-1\right)^{ab}J^b_R(y)\right\}}
\end{equation}
with
\begin{equation}
J_V^a(x)=J_R^a(x)-J_L^a(x)
\end{equation}
The operators $J_V^a(x)$ are the generators of $SU_V(N_c)$ - the vector subgroup of $SU_L(N_c)\otimes SU_R(N_c)$.

As noted above, the JIMWLK Hamiltonian is obtained from eq.(\ref{klwmij}) by the dense-dilute duality transformation
\begin{equation}\label{densedilute}
R(x)\rightarrow S(x)
\end{equation}
We note that  a generalization of $H_{RFT}$ which interpolates between $H_{KLWMIJ}$ and $H_{JIMWLK}$ has recently been derived \cite{aklp}. However due to its complexity we will not deal with it in the present paper.

The Hamiltonian $H_{KLWMIJ}$ is the limit of $H_{RFT}$ at low color charge density. The left and right rotation operators $J_R$ and $J_L$ eq.(\ref{LR}) that appear in eq.(\ref{klwmij}) are in fact just the color charge density in the hadronic wave function and its conjugate respectively \cite{kl}. 

 Some aspects of the structure of the spectrum of $H_{KLWMIJ}$ were discussed in \cite{yinyang}. In particular we know that $H_{KLWMIJ}$ is positive definite since it can be written as
\begin{equation}
H_{KLWMIJ}=\frac{\alpha}{2\pi^2}\int_zQ^{a \dagger}_i(z)Q^a_i(z)
\end{equation}
with the Hermitian amplitude 
\begin{equation}
 Q^a_i(z)=\int_x\frac{(x-z)_i}{(x-z)^2}\left[R^{ab}(z)-R^{ab}(x)\right]J^b_R(x) 
\end{equation}
It has two states with zero eigenvalue:
\begin{equation}
H_{KLWMIJ}|Yin\rangle=0; \ \ \ \ \ \ H_{KLWMIJ}|Yang\rangle=0
\end{equation}
The state $|Yang\rangle$ corresponds to the physical vacuum, that is to the state which is annihilated by the color charge density
\begin{equation}
J^a_L(x)|Yang\rangle=J^a_R(x)|Yang\rangle=0
\end{equation}
while $|Yin\rangle$ represents the black disk state
\begin{equation}
R^{ab}(x)|Yin\rangle=\delta^{ab}|Yin\rangle
\end{equation}
We will find it convenient to work in the basis of eigefunctions of the operator $R$.
Written in this basis the wave functionals of the two states are
\begin{equation}
\langle R|Yang\rangle=1; \ \ \ \ \ \ \langle R|Yin\rangle=\delta\left(R^{ab}(x)-\delta^{ab}\right)
\end{equation}
Each one of these states sustains a tower of excitations above it. The RFT states "close" to $|Yang\rangle$ correspond to physical QCD states with small number of particles in the projectile wave function. This interpretation stems from the fact that, as discussed in detail in \cite{kl} the probability distribution $W$ has a convenient representation
\begin{equation}
W[\rho]=\Sigma[R]\delta[\rho]
\end{equation}
and thus in the $R$ basis $W$ is simply a regular function of $R$. Every factor of $R$ in $W$ corresponds to a gluon in the projectile wave function. Thus expansion of $W$ in powers of $R-1$, is equivalent to expansion in the number of the projectile gluons that scatter on the target.

In this paper we are interested in the eigenstates of $H_{KLWMIJ}$ which have similar structure, namely in the $R$ basis the eigenfunctions are regular functionals of $R(x)$ which can be expanded in powers of $R-1$. Since $R$ is a regular function of $\delta/\delta\rho$, these same states can also be expanded in powers of $\delta/\delta\rho$. Clearly the state $|Yang\rangle$ is one of those states since the wave function in this case is simply a constant. 

On the other hand the state $|Yin\rangle$ does not fall into this category. Its wave function is not expandable in powers of $R-1$. The same is true for other states "close" to it. As explained in \cite{yinyang} those states correspond to "holes" in the black disk and their eigenfunctions are expandable in powers of $\rho$ rather than $\delta/\delta \rho$.

The discussion of this paper pertains directly only to the $|Yang\rangle$ - like states.

The derivation of the Hamiltonian $H_{KLWMIJ}$ \cite{kl} does not assume anything about the strength of the target fields. If one interested in the situation when the target fields are small, one can expand $H_{KLWMIJ}$  in powers of $\delta/\delta\rho$. The expansion in powers of $\delta/\delta\rho$ is equivalent to expansion in powers of the target field, since powers of $\delta/\delta\rho$ turn into powers of $\alpha_T$ in the calculation of the scattering matrix eq.(\ref{smatrix1}). The leading order of this expansion gives $H_{BFKL}$, which is the second quantized Hamiltonian that generates the high energy evolution in the BFKL framework.
The form of $H_{BFKL}$ is well known (see for example \cite{msw}). For completeness we present the derivation
of $H_{BFKL}$ in the appendix, carefully keeping track of the path ordering in the definition of $R(x)$. Although this path ordering is not relevant in many cases \cite{reggeon}, in general it cannot be neglected. The result is 
\begin{equation}\label{bfkl}
H_{BFKL}=-\frac{\alpha_s}{2\pi^2}\int_{xyz}K_{xyz}(T^aT^b)_{cd}\rho^a_x\left[\frac{\delta}{\delta \rho^c_x}-\frac{\delta}{\delta \rho^c_z}\right]\left[\frac{\delta}{\delta \rho^d_y}-\frac{\delta}{\delta \rho^d_z}\right]\rho^b_y
\end{equation}
where
\begin{equation}
\rho^a_x\equiv\int_0^1dx^-\rho^a(x,x^-); \ \ \ \ \ \frac{\delta}{\delta\rho^a_x}\equiv\int_0^1dx^-\frac{\delta}{\delta\rho^a(x,x^-)}
\end{equation}
In the "normal ordered form" this reads
\begin{equation}\label{bfklno}
H_{BFKL}=-\frac{\alpha_s}{2\pi^2}\int_{xyz}K_{xyz}(T^aT^b)_{cd}\left[\frac{\delta}{\delta \rho^c_x}-\frac{\delta}{\delta \rho^c_z}\right]\left[\frac{\delta}{\delta \rho^d_y}-\frac{\delta}{\delta \rho^d_z}\right]\rho^a_x\rho^b_y+\int_{xz}\beta_{xz}\frac{\delta}{\delta \rho^a_z}\rho^a_x
\end{equation}
with
\begin{equation}\label{beta}
\beta_{x-z}=\frac{\alpha_sN_c}{2\pi^2}\left[\delta^2(x-z)\int_uK(x,x,u)-K(x,x,z)\right]
=\frac{\alpha_sN_c}{ 2\pi^2}\left[\delta^2(x-z)\int_u\frac{1}{u^2}-\frac{1}{(x-z)^2}\right]
\end{equation}

The question we want to address is what is the relation between the eigenvalues and eigenfunctions of $H_{BFKL}$ and $H_{KLWMIJ}$. More importantly, to what extent can we use the results of calculations performed in the framework of BFKL evolution to get information about the spectrum of $H_{KLWMIJ}$.

\section{From BFKL to KLWMIJ}
The calculation of the spectrum of $H_{BFKL}$ is equivalent to solution of the complete set of BKP equations.

The BFKL Hamiltonian is a homogeneous function of the coordinates ($\delta/\delta \rho$) and momenta ($\rho$) and thus its eigenfunctions are pure powers. Taking an eigenfunction in the form
\begin{equation}
\Psi_A[\delta/\delta\rho]=\int_{x_1...x_n}G_A^{a_1a_2...a_n}(x_1...x_n)\frac{\delta}{\delta\rho^{a_1}_{x_1}}...\frac{\delta}{\delta\rho^{a_n}_{x_n}}
\end{equation}
and acting on it with $H_{BFKL}$ leads to an eigenvalue equation of the form
\begin{eqnarray}\label{bkp}
&&-\frac{\alpha}{2\pi^2}\Sigma_{i\ne j}(T^aT^b)_{a_ia_j}\Bigg[\int_zK_{x_ix_jz}G_A^{a_1...a...b...a_n}(x_1...x_n)+\int_{xy}K_{xyx_i}G_A^{a_1...a...b...a_n}(x_1...x...y...x_n)\delta(x_i-x_j)\nonumber\\
&&-\int_yK_{x_iyx_j}G_A^{a_1...a...b...a_n}(x_1...x_i...y...x_n)-\int_xK_{xx_jx_i}G_A^{a_1...a...b...a_n}(x_1...x...x_j...x_n)\Bigg]\nonumber\\
&&+\Sigma_i\int_x\beta_{xx_i}G_A^{a_1...a...b...a_n}(x_1...x...x_n)=\omega_AG_A^{a_1...a_n}(x_1...x_n)
\end{eqnarray}
Eqs.(\ref{bkp}) are precisely the BKP equations \cite{bkp} for the rapidity evolution of $n$ - gluon exchange in $t$-channel.
The index $A$ denotes various quantum numbers that characterize the eigenfunction $G_A$, in particular total momentum, color representation, charge conjugation, parity and so on.

As mentioned above, expanding $H_{KLWMIJ}$ and $\Psi$ is powers of $\delta/\delta\rho$ is equivalent to expanding the $S$ matrix eq.(\ref{smatrix}) in powers of the target color field $\alpha_T$. Every factor of $\alpha_T$ represents an exchange of a gluon in $t$-channel between the projectile and the target. Thus physically, expansion in powers of $\delta/\delta\rho$ is equivalent to expansion in the number of gluons exchanged in the $t$ - channel \cite{reggeon}. Consequently, the quantum numbers denoted by the index $A$ in eq.(\ref{bkp}) are the quantum numbers of the $n$ - gluon state exchanged in the $t$-channel.

 For $n=1$ the solution of eq.(\ref{bkp}) is the reggeized gluon. In this case the color representation of the exchange is obviously adjoint and the eigenvalues are characterized by the transverse momentum. For the singlet two gluon state, $n=2$, this is the celebrated BFKL equation and the solution is the BFKL Pomeron.  For arbitrary $n$ in the large $N_c$ approximation these equations have been extensively studied in \cite{korchemsky} where it was shown that the spectrum of the $n$ gluon state is described by an integrable spin chain.

Although the full spectrum of eq.(\ref{bkp}) is not known, the salient features are the following. In the reggeized gluon sector ($n=1$) the eigenvalues are nonnegative. The lowest eigenvalue is vanishing and corresponds to zero transverse momentum exchange $t$. At nonzero $t$ the eigenvalues are logarithmically infrared divergent. For the BFKL Pomeron (singlet $n=2$ exchange) the lowest eigenvalue is actually negative, corresponding to the growth of the amplitude at high energy with the famous BFKL intercept. At $n>2$ the spectrum is rich. Importantly the lowest eigenvalue for the color singlet exchange is always negative and grows proportionally to $n$, corresponding to the $n/2$ Pomeron exchanges \footnote{In fact for very large $n>N_c$ this growth is even faster and the absolute value of the most negative eigenvalue is proportional to $n^2$ \cite{LALE}.} 

Now let us consider $H_{KLWMIJ}$. The eigenvalues and eigenfunctions as noted above are determined by solving
\begin{equation}
H_{KLWMIJ}\Psi_A[R]=\omega_A\Psi_A[R]
\end{equation}
Suppose we try to solve this by Taylor expanding $\Psi_A$ in powers of $\delta/\delta\rho$. We thus write
\begin{equation}\label{expfu}
\Psi_A=\Psi^0_A[\frac{\delta}{\delta\rho}]+\Psi^1_A[\frac{\delta}{\delta\rho}]+...
\end{equation}
To find $\Psi$ we need to act on it with $H_{KLWMIJ}$ also expanded in powers of $\delta/\delta\rho$.
It is clear from the mechanics of the expansion of $J_L$ and $J_R$ given in the appendix, that to all orders in $\delta/\delta\rho$, both the charge densities are proportional to the first power of $\rho$, and only the power of $\delta/\delta\rho$ in $H_{KLWMIJ}$ grow order by order. Therefore the Hamiltonian $H_{KLWMIJ}$ can be written as
\begin{equation}\label{expeq}
H_{KLWMIJ}=H_{BFKL}+H_1+...
\end{equation}
Here the first correction to the Hamiltonina, $H_1$ schematically has the form
\begin{equation}
H_1=K_1\left(\frac{\delta}{\delta\rho}\right)^3\rho^2+\beta_1\left(\frac{\delta}{\delta\rho}\right)^2\rho
\end{equation}
The structure of eqs.(\ref{expfu},\ref{expeq}) is such that in the leading order the wave function $\Psi^0_A[\frac{\delta}{\delta\rho}]$ must be an eigenfunction of $H_{BFKL}$. Thus the leading order equation determines $\omega_A$ completely, and the role of the higher order equations is only to determine the higher order Taylor series terms in the expansion of the wave function $\Psi_A[R]$.
E.g. having found $\Psi^0_A$ as an eigenfunction of $H_{BFKL}$ with eigenvalue $\omega_A$, to first order one has
\begin{equation}
\Psi^1_A[\rho]=\frac{1}{ \omega_A-H_{BFKL}}H_1\Psi^0_A
\end{equation}
and so on.

It thus appears that the eigenvalues of $H_{KLWMIJ}$ can be obtained {\it exactly} from the leading order approximation - the BFKL Hamiltonian. The catch however is that we do not know from the BFKL calculation {\it per se} whether the wave function corresponding to a given "eigenvalue" will turn out to be normalizable or not. Within the BFKL approximation itself, none of the wavefunctions are of course normalizable since they are simple monomials of $\delta/\delta\rho$. However only normalizable eigenfunctions of $H_{KLWMIJ}$ should be used in the expansion of the $S$-matrix eqs.(\ref{psii},\ref{psiii}).

A similar situation is encountered in simple quantum mechanics. Take for example a one dimensional harmonic oscillator 
\begin{equation}
h=\frac{1}{2}\left(p^2+x^2\right)
\end{equation}
Let us now try to find its ground state wavefunction in Taylor expansion. We take
\begin{equation}
\psi=1+ax^2+cx^4+...
\end{equation}
Acting on it with the Hamiltonian we get
\begin{equation}
h\psi=-a +\frac{1}{2}x^2-6cx^2...
\end{equation} 
and the Schroedinger equation
\begin{equation}
-a+\frac{1}{2}(1-12c)x^2=\omega(1+ax^2)
\end{equation}
Thus for arbitrary $\omega$ we simply have
\begin{equation}
a=-\omega; \ \ \ \ \ c=\frac{1}{12}(1+2\omega^2)
\end{equation}
So there is a solution for every possible value of $\omega$ (in fact there are two solutions for each $\omega$, but the other one is an odd function of $x$ and is outside our initial ansatz).
This calculation however does not carry a lot of information since we do not know {\it a priori} which of the so found functions are normalizable when the Taylor series is summed to all orders. We know of course that the spectrum in fact is discreet and therefore most of the "eigenvalues" do not correspond to normalizable eigenfunctions.

The situation in the KLWMIJ-BFKL system is similar in this respect. As we have noted above, $H_{KLWMIJ}$ is hermitian and positive definite, thus its eigenvalues have to be positive. On the other hand many of the eigenvalues of $H_{BFKL}$ are negative (including of course, the Pomeron). We are therefore assured that those eigenvalues do not correspond to normalizable eigenfunctions. As for the positive eigenvalues of $H_{BFKL}$, it is tempting to surmise that they correspond to normalizable eigenfunctions of $H_{KLWMIJ}$. Unfortunately, we have no right to do so. The question whether resummed Taylor series is normalizable or not is very complicated and can not be answered in any finite order of Taylor expansion. We note however that the Taylor expansion for $H_{KLWMIJ}$ is in a subtle way different from that for the harmonic oscillator. In the later case the leading order of the expansion puts no restrictions at all on possible eigenvalues. In the KLWMIJ case however, the leading order itself leads to an eigenvalue problem, so that not every $\omega_A$ is allowed. 

Even though it is not clear whether the eigenfunctions of $H_{BFKL}$ give rise to normalizable eigenfunctions of $H_{KLWMIJ}$, it is still interesting to illustrate the procedure discussed above by some concrete example. In the following section we will therefore consider higher order in $\delta/\delta\rho$ corrections to some eigenfunctions. We will concentrate on the eigenvalues corresponding to the reggeized gluon, which is the simplest eigenfunction of $H_{BFKL}$. 

\section{The Reggeized gluon and the bootstrap.}
\subsection{The Reggeized gluon}
Let us look for the eigenstate of $H_{KLWMIJ}$ which corresponds to the quantum numbers of one gluon exchange. The state must belong to the adjoint representation of $SU_V(N_c)$
and its wave function when Taylor expanded should start with the linear term in $\delta/\delta\rho$. 
Thus we take 
\begin{equation}\label{regg}
\Psi_0=\int d^2x \phi(x)\frac{\delta}{\delta\rho^a(x)}
\end{equation}
Acting on it with $H_{BFKL}$
we obtain the eigenvalue equation
\begin{equation}\label{omegabeta}
\int_z\beta_{xz}\phi_z=\omega\phi_x
\end{equation}
This is solved by
\begin{equation}\label{eigenf}
\phi(x)=e^{iqx}
\end{equation}
with the eigenvalue
\begin{equation}\label{intercept}
\omega_q=\frac{\bar{\alpha}}{2\pi}\int_\mu d^2k\frac{q^2}{k^2(q-k)^2}
\end{equation}
This is nothing but the gluon reggeization. 

To calculate first correction to the reggeized gluon state we have to consider $\Psi^1$ which is quadratic in $\delta/\delta \rho$. 
We will perform the calculation in a slightly different way which is technically simpler. We know that the wave function $\Psi$ at the end of the day should depend only on $R$. Therefore it makes sense, rather than taking an arbitrary quadratic function, to choose such a function that itself can be obtained from expansion of $R$.
Since $\Psi^0$ is simply represented as expansion of $R$ 
\begin{equation}
\Psi^a_0[\delta/\delta \rho]=\int d^2x\phi(x)tr \left(T^aR\right)_{\rm first \ order}
\end{equation}
we will take a simple guess
\begin{equation}
\Psi^a_0[\delta/\delta \rho]+\Psi_a^1[\delta/\delta \rho]=\int d^2x\phi(x)tr \left(T^aR\right)_{\rm first+second \ order}
\end{equation}
and will show that it indeed satisfies the KLWMIJ eigenvalue equation to second order in $\delta/\delta \rho$. 
The matrix $R$ is taken here in the adjoint representation.

\subsection{Bootstrap of the antisymmetric adjoint}
%We will take the KLWMIJ Hamiltonian in the form
%\begin{equation}
%H_{KLWMIJ}=K_{xyz}\left\{2J^c_L(x)[R(z)-1]^{cd}J^d_R(y)-[J^c_L(x)-J^c_R(x)][J^c_L(y)-J^c_R(y)]\right\}
%\end{equation}
%and one gluon state is
We now consider the action of $H_{KLWMIJ}$ in the state 
\begin{equation}
G^a_A= \int d^2x\phi(x)tr \left(T^aR\right)
\end{equation}
The basic elements we need is the action of the left and right rotation generators on the matrix $R$
\begin{equation}
[J^a_L(x),R(y)]=T^aR(y)\delta^2(x-y); \ \ \ \ \ \ [J^a_R(x),R(y)]=R(y)T^a\delta^2(x-y)
\end{equation}
With this it is easy to calculate the action of the real and virtual parts of $H_{KLWMIJ}$:
\begin{equation}
2J^c_L(x)[R(z)-1]^{cd}J^d_R(y)tr[T^aR_u]=2[R_z-1]^{cd} tr[T^aT^cR_xT^d]\delta(x-y)\delta(y-u)
\end{equation}
and 
\begin{equation}
[J^c_L(x)-J^c_R(x)][J^c_L(y)-J^c_R(y)]tr[T^aR_u]=tr\left\{T^a[T^c,[T^c,R_x]]\right\}\delta(x-y)\delta(y-u)
\end{equation}
Hence 
\begin{equation}
H_{KLWMIJ}G^a_A=-\int_{u,z}K_{uuz}\phi_u\left\{2[R_z-1]^{cd}tr[T^dT^aT^cR_u]+tr\left\{T^a[T^c,[T^c,R_u]]\right\}\right\} 
\end{equation}
Using
\begin{equation}
tr\left\{T^a[T^c,[T^c,R_u]]\right\}=N_ctr[T^aR_u]
\end{equation}
we write
\begin{eqnarray}\label{1glue}
H_{KLWMIJ}G^a_A&=&-\frac{\alpha}{ 2\pi^2}\int_{u,z}K_{uuz}\phi_u\left\{2[R_z-1]^{cd}tr[T^dT^aT^cR_u]-N_ctr[T^aR_u]\right\} \nonumber \\
&=&\frac{\alpha}{ \pi^2}\int_{u,z}K_{uuz}\phi_u\left\{-2[R_z-1]^{cd}tr[T^dT^aT^c(R_u-1)]+N_c\left(tr[T^aR_u]-tr[T^aR_z]\right)\right\}
\end{eqnarray}
We now have to expand this to second order in $\delta/\delta\rho$. 
The matrix $R$ is expanded as
\begin{equation}
R^{ab}_u=\delta^{ab}+T^{ab}_c\int^1_0du^-\frac{\delta}{\delta \rho^c(u,u^-)}+T^{a e}_c T^{e b}_d \int^1_0du^-_1\int^{u^-_1}_0du^-_2\frac{\delta}{\delta \rho^c(u,u^-_1)}\frac{\delta}{\delta \rho^d(u,u^-_2)}
\end{equation}
The crucial observation is that the first and second order terms come only from the second term in eq.(\ref{1glue}). For first order term this is obvious. The second order contribution from the first term in eq.(\ref{1glue}) is proportional to
\begin{equation}
T^e_{cd}tr[T^dT^aT^cT^b]\frac{\delta}{\delta\rho_u^b}\frac{\delta}{\delta\rho_z^e}
\end{equation}
However, using the properties of the adjoint SU(N) generators
\begin{equation}
tr(T^aT^b)=f^{a\alpha\beta}f^{b\alpha\beta}=N_c\delta^{ab} ,\ \ tr(T^aT^b T^c)=-if^{\alpha a\beta}f^{\beta b\gamma}f^{\gamma c\alpha}=i\frac{N_c}{2}f^{abc}
\end{equation}
one can easily show that 
\begin{equation}
T^e_{cd}tr[T^dT^aT^cT^b]=0
\end{equation}
and so this contribution vanishes.
Thus to second order the eigenvalue equation is
\begin{eqnarray}\label{antisreg}
&&\frac{\alpha N_c}{2\pi^2}\int_{u,z}K_{uuz}\phi_u\left[\left(\frac{\delta}{\delta\rho_u^a}-\frac{\delta}{\delta\rho_z^a}\right)+\frac{i}{2} f^{abc} \int_{x^->y^-} \left(\frac{\delta}{\delta\rho^b(u,x^-)}\frac{\delta}{ \delta\rho^c(u,y^-)}-\frac{\delta} {\delta\rho^b(z,x^-)}\frac{\delta}{ \delta\rho^c(z,y^-)}\right)\right]\nonumber\\
&&=
\omega\int_u\phi(u)\left[\frac{\delta}{\delta\rho_u^a}+\frac{i}{ 2}f^{abc} \int_{x^->y^-} \frac{\delta}{ \delta\rho^b(u,x^-)}\frac{\delta}{ \delta\rho^c(u,y^-)}\right]
\end{eqnarray}
Obviously, this is satisfied as before by $\phi(x)$ of eq.(\ref{eigenf}) with the eigenvalue eq.(\ref{intercept}).
According to our earlier discussion, the eigenvalue determined in the leading order does not change order by order. An interesting feature of this calculation is that the second order correction to the eigenfunction is the term which describes the two gluon exchange in the $t$-channel such that the two gluons are in the octet. The fact that this term reggeizes precisely in the same way as the one gluon exchange is the essence of the celebrated bootstrap feature in high energy QCD \cite{bootstrap}. 

\subsection{Bootstrap of the symmetric adjoint}
The two $t$-channel gluons in the previous calculation are in the antisymmetric adjoint representation. It is easy to show that two gluons in the symmetric adjoint also reggeize (see second paper in \cite{reggeization}).
To see this let us consider a similar calculation but take the matrix $R$ to be in the fundamental representation. 
As before we take
\begin{equation}\label{quark}
G^a_F=\int_u\phi_utr[\tau^aR_F(u)]
\end{equation}
where $\tau^a$ are generators of $SU(N_c)$ in the fundamental representation.
The action of $H_{KLWMIJ}$ on this state is claculated just like before using
\begin{equation}
[J^a_L(x),R_F(y)]=\tau^aR_F(y)\delta^2(x-y); \ \ \ \ \ \ [J^a_R(x),R_F(y)]=R_F(y)\tau^a\delta^2(x-y)
\end{equation}

Using of completeness relation of fundamental SU(N) generators
\begin{equation}
\tau^c_{\alpha \beta}\tau^c_{\gamma \delta}=\frac{1}{2}\left[\delta_{\alpha \delta}\delta_{\beta \gamma}-\frac{1}{N_c}\delta_{\alpha \beta}\delta_{\gamma \delta}\right]
\end{equation}
the properties of fundamental generators
\begin{equation}
tr(\tau^a\tau^b)=\frac{1}{2}\delta^{a b}, \  \ tr(T^aT^b T^c)=\frac{1}{4}(d^{abc}+if^{abc})
\end{equation}
and the representation of an adjoint unitary matrix in terms of fundamental matrices
\begin{equation}
R^{ab}_A(z)=2tr\left[\tau^aR_F(z)\tau^bR^{\dagger}_F(z)\right]
\end{equation}
one can write
%\begin{equation}
%T^c_{\alpha \beta}T^c_{\beta \gamma}=\frac{1}{2}\left[\delta_{\alpha \gamma}\delta_{\beta \beta}-\frac{1}{N}\delta_{\alpha \beta}\delta_{\beta \gamma}\right]=\frac{N^2-1}{2N}\delta_{\alpha \gamma}
%\end{equation}
%and
%\begin{eqnarray}
%tr\left[T^aT^cR_uT^c\right]&=&T^a_{\alpha \beta}T^c_{\beta \gamma}(R_u)_{\gamma \lambda}T^c_{\lambda \alpha} \nonumber \\
%&=&\frac{1}{2}\left[\delta_{\beta \alpha}\delta_{\gamma \lambda}-\frac{1}{N}\delta_{\beta \gamma}\delta_{\lambda \alpha}\right]T^a_{\alpha \beta}(R_u)_{\gamma \lambda} \nonumber \\
%&=&-\frac{1}{2N}tr[T^aR_u]
%\end{eqnarray}
%So
%\begin{equation}
%tr\left\{T^a\left[T^c,\left[T^c,R_x\right]\right]\right\}=Ntr[T^aR_u]
%\end{equation}
%Now consider the first term
%\begin{equation}
%2\left[R_A(z)-1\right]^{cd}tr\left[T^aT^cR_uT^d\right]=2R_A^{cd}(z)tr\left[T^aT^cR_uT^d\right]-2\delta^{cd}tr\left[T^aT^cR_uT^d\right]
%\end{equation}
%We already calculated
%\begin{equation}
%2\delta^{cd}tr\left[T^aT^cR_uT^d\right]=-\frac{1}{N}tr[T^aR_u]
%\end{equation}
%By using the identity that relates adjoint and fundamental representations of a unitary matrix
%one can write
%\begin{eqnarray}
%2R^{cd}_A(z)tr\left[T^aT^cR_uT^d\right]&=&4tr\left[T^aR_zT^cR^{\dagger}_z\right]tr\left[T^aT^cR_uT^d\right]\nonumber \\
%&=&tr[R^{\dagger}_zR_u]tr[T^aR_z]-\frac{1}{N}tr[T^aR_u]
%\end{eqnarray}
%So
%\begin{equation}
%2\left[R_A(z)-1\right]^{cd}tr\left[T^aT^cR_uT^d\right]=tr[R^{\dagger}_zR_u]tr[T^aR_z]
%\end{equation}
%With all these information we can write KLWMIJ evolution of one quark state as
\begin{equation}\label{regf}
H_{KLWMIJ}G^a_F=\frac{\alpha}{ 2\pi^2}\int_{u,z}K_{uuz}\phi_u\left\{tr[1-R^{\dagger}_{F}(z)R_{F}(u)]tr[\tau^aR_{F}(z)]+N_ctr\left[\tau^a\left(R_{F}(u)-R_{F}(z)\right)\right]\right\}
\end{equation}
The first term on the RHS starts in the order $(\delta/\delta\rho)^3$. Thus the eigenvalue equation to second order reads
\begin{eqnarray}\label{symreg}
&&\frac{\alpha N_c}{ 2\pi^2}\int_{u,z}K_{uuz}\phi_u\left[\left(\frac{\delta}{\delta\rho_u^a}-\frac{\delta}{\delta\rho_z^a}\right)+\frac{1}{ 2} \left\{if^{abc}+d^{abc}\right\} \int_{x^->y^-} \left(\frac{\delta}{ \delta\rho^b(u,x^-)}\frac{\delta}{ \delta\rho^c(u,y^-)}-\frac{\delta}{ \delta\rho^b(z,x^-)}\frac{\delta}{ \delta\rho^c(z,y^-)}\right)\right]\nonumber\\
&&=
\omega\int_u\phi(u)\left[\frac{\delta}{\delta\rho_u^a}+\frac{1}{ 2} \left\{if^{abc}+d^{abc}\right\} \int_{x^->y^-} \frac{\delta}{ \delta\rho^b(u,x^-)}\frac{\delta}{ \delta\rho^c(u,y^-)}\right]
\end{eqnarray}
Again the plane wave eq.(\ref{eigenf}) is the solution of this equation with the eigenvalue eq.(\ref{intercept}). The second order term proportional to the $d^{abc}$ tensor corresponds to exchange of two $t$ - channel gluons in the symmetric adjoint representation. As we have seen in the previous subsection, the antisymmetric octet (the $f^{abc}$ term) reggeizes. Thus eq.(\ref{symreg}) tells us that the symmetric adjoint reggeizes by itself. This is another example of bootstrap at work.

\subsection{The bootstrap in the KLWMIJ/JIMWLK approach.}
Since the bootstrap plays such an important role in the discussions of high energy amplitudes, it is worth while explaining how and why the bootstrap condition  in the KLWMIJ approach is satisfied {\it by fiat}.

First off, we note that the bootstrap condition, which leads to reggeization of the two gluon exchange can be stated as the relation between the real part of the kernel of the BFKL equation and the reggeized gluon trajectory\cite{levin}.
\begin{equation}
\omega(k_1)-\omega(k_2)-\omega(k_1+k_2)=\frac{1}{ 2}\int_{q_1,q_2}\tilde K(k_1,k_2,q_1,q_2)
\end{equation}
Here $\tilde K$ is the real part of the BFKL kernel which evolves the color singlet two $t$-channel gluon state. It arises from the first term in $H_{BFKL}$ eq.(\ref{bfklno}). Expressing it in coordinate space in terms of the kernel $K(x,y,z)$ we have
%\begin{equation}
%\tilde 2\int_z\left[K_{xyz}\psi(x,y)-K_{xzy}\psi(x,z)-K_{zyx}\psi(z,y)+\delta(x-y)K_{zux}\psi(z,u)\right]
%\end{equation}
%One can write this expression as a function of variables in the following way
\begin{equation}
\tilde K(xy,uv)=2\Bigg[\int_zK_{uvz}\delta(x-u)\delta(y-v)-K_{uvy}\delta(x-u)
-K_{uvx}\delta(y-v)+K_{uvx}\delta(x-y)\Bigg]
\end{equation}
%In the Fourier space 
%\begin{equation}
%\psi(k_1,k_2)=\int_{x,y}\psi(x,y)e^{ik_1x+ik_2y}
%\end{equation}
%K is the real part of the momentum space BFKL kernel
%\begin{equation}
%K(k_1,k_2;q_1,q_2)=\frac{\alpha_s}{2\pi}\left[\frac{(q_1-k_1)_i}{(q_1-k_1)^2}-\frac{q_{1i}}{q_1^2}\right]\left[\frac{(q_2-k_2)_i}{(q_2-k_2)^2}-\frac{q_{2i}}{q_2^2}\right]\delta(k_1+k_2-q_1-q_2)
%\end{equation}
According to eq.(\ref{omegabeta}) and eq.(\ref{eigenf}) the Fourier transform of $\omega(k)$ into coordinate space is $\beta(x)$.
Fourier transforming the bootstrap condition into the coordinate space we have
\begin{equation}\label{configbootstrap}
\beta(x)\delta(y)-\beta(y)\delta(x)-\beta(x)\delta(x-y)=\Bigg[\int_zK_{00z}\delta(x)\delta(y)-K_{00y}\delta(x)-K_{00x}\delta(y)+K_{00x}\delta(x-y)\Bigg]
\end{equation}
We can derive this condition directly by considering the eigenvalue equation in the symmetric adjoint channel.
Take the trial function in the form
\begin{equation}
G^a=\int_{uv}\Psi(u,v)d^{abc}\frac{\delta}{\delta \rho^b_u}\frac{\delta}{\delta \rho^c_v}
\end{equation}
Acting on it by $H_{BFKL}$ we derive the eigenvalue equation
\begin{eqnarray}
-\frac{N_c\alpha}{4\pi^2}\int\Big[2K_{uvz}\Psi(u,v)-2K_{uzv}\Psi(u,z)-2K_{zvu}\Psi(z,v)+2K_{zxu}\delta(u-v)\Psi(z,x)\Big]&+&\Big[\beta_{zv}\Psi(z,u)+\beta_{zu}\Psi(z,v)\Big]\nonumber \\
&=&\omega_q\Psi(u,v)
\end{eqnarray}
Assuming that the solution identical to the single gluon exchange
\begin{equation}
\Psi_q(u,v)=\delta^2(u-v)e^{iqu}
\end{equation}
exists for all $q$, and taking the integral over $q$ we arrive at the configuration space condition eq.(\ref{configbootstrap}) \footnote{We note that the condition Eq.(\ref{configbootstrap}) can not be derived by acting with the BFKL Hamiltonian on the color octet antisymmetric two gluon state, even though this state does reggeize. The reason is that the reggeization of the antisymmetric octet is technically somewhat different form the symmetric one. The antisymmetric state $\Psi^A_2$ appears in the wave function $\Psi$ in a combination with the one gluon state $\Psi_1$. Contributions to the two gluon antisymmetric term in the left hand side of the eigenvalue equation eq.(\ref{antisreg}) arise both from the action of $H_{BFKL}$ on $\Psi^A_2$ {\bf and} from the action of $H_1$ on $\Psi_1$. On the other hand the action of $H_1$ on $\Psi_1$ does not generate any contributions to the symmetric octet function. Thus the symmetric octet reggeizes  by the action of $H_{BFKL}$ alone, but the antisymmetric state does not.}.

With $\beta(x)$ defined in eq.(\ref{beta}), this condition is clearly satisfied. Clearly, the bootstrap condition would be satisfied in the KLWMIJ approach for any functional form of kernel $K_{xyz}$, since the relation between $\beta$ and $K$, eq.(\ref{beta}) is immutable. It appears simply due to normal ordering of $H_{BFKL}$ written in the original form eq.(\ref{bfkl}). Thus even if $K$ is modified in eq.(\ref{bfkl}), the bootstrap condition will still be automatically satisfied. One can contemplate several reasons for such a modification. First, higher order corrections in $\alpha_s$ certainly lead to modification of $K$ \cite{nexttoleading}. Another reason to consider a modification of $K$ is the unphysical infrared behavior of perturbative gluon emission which leads to violation of Froissart bound \cite{froissart}. Cutting off long distance tails of the Weiszacker-Williams field in the emission kernel $K$ is a possible "phenomenological" solution of this problem \cite{cutoff}.
One could question in principle the starting point of our discussion - eq.(\ref{bfkl}).
However the form eq.(\ref{bfkl}) is dictated by the hermiticity of $H_{BFKL}$. The real emission part is given by the first term in eq.(\ref{bfklno}). By itself this term is not hermitian, and only with the gluon trajectory term, the second term in eq.(\ref{bfklno}), the hermiticity of the Hamiltonian is restored. 

Thus we conclude that the bootstrap condition within the KLWMIJ framework is tantamount to the condition of hermiticity of the Hamiltonian which generates the rapidity evolution of the scattering amplitude.

The Hermiticity of the Hamiltonian is related to the unitarity of the high energy evolution, even though the evolution equation is not a Schroedinger equation, but rather a diffusion type equation. The evolution Hamiltonian acts on the probability distribution $W$ eq.(\ref{evoleq}). Since $W[\rho]$ has the meaning of probability density, it must be positive definite. The eigenvalues of the evolution thus better be real, otherwise the "probability density" will develop an imaginary part even if one starts with a real and positive distribution at initial rapidity. This is assured if the evolution Hamiltonian is Hermitian.

 The origin of reggeization,  including the gluon rerggeization, is in the $t$-channel unitarity. On the other hand the JIMWLK/KLWMIJ approach  is formulated in  $s$-channel and has the most natural interpretation as the evolution of the $s$-channel wave function. It is therefore interesting to see how $t$ and $s$ channel pictures are interrelated on the example of the gluon reggeization.  The two pictures lead to the equivalent description, if the evolution in rapidity can be described by a hermitian Hamiltonian.

 We note that the bootstrap equation was used in \cite{levin} to find the generalization of  the gluon reggeization for the running QCD coupling case.
 The JIMWLK/KLWMIJ Hamiltonian is of course modified when the running is taken into account.
The practical conclusion from our discussion in this subsection is the following: if we know how to include the running of the QCD coupling in $\beta$ of  \eq{beta}, then \eq{bfklno} can be used to generalize the full BFKL Hamiltonian to the running coupling case.  This idea has been explored in \cite{levin} and led to the "triumvirate" structure \cite{KOVWE}. The same form of the Hamiltonian was derived recently in \cite{KOVWE} by direct summation of the Feyman diagrams in the dipole approximation. 
\eq{bfklno} with 
\begin{equation}\label{interceptf}
\omega_q=\frac{1}{2\pi}\int_\mu d^2k\,\frac{\bar{\alpha}(\left( (q  - k)^2\right) \,\bar{\alpha}\left( k^2\right) }{
\bar{\alpha}\left(q^2\right)}\frac{q^2}{k^2(q-k)^2}
\end{equation}
gives the correct generalization of the JIMWLK/KLWMIJ Hamiltotian for running QCD coupling both for linear and non-linear term at any value of $N_c$.

\section{Screening corrections}

\begin{figure}
\includegraphics{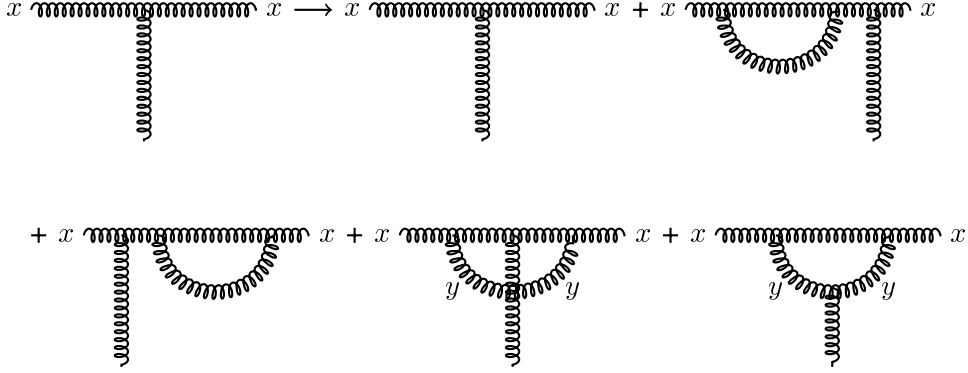}
\caption{\label{f1} KLWMIJ evolution in the approximation where only one gluon exchange in the $t$-channel is allowed.}
\end{figure}

\begin{figure}
\includegraphics{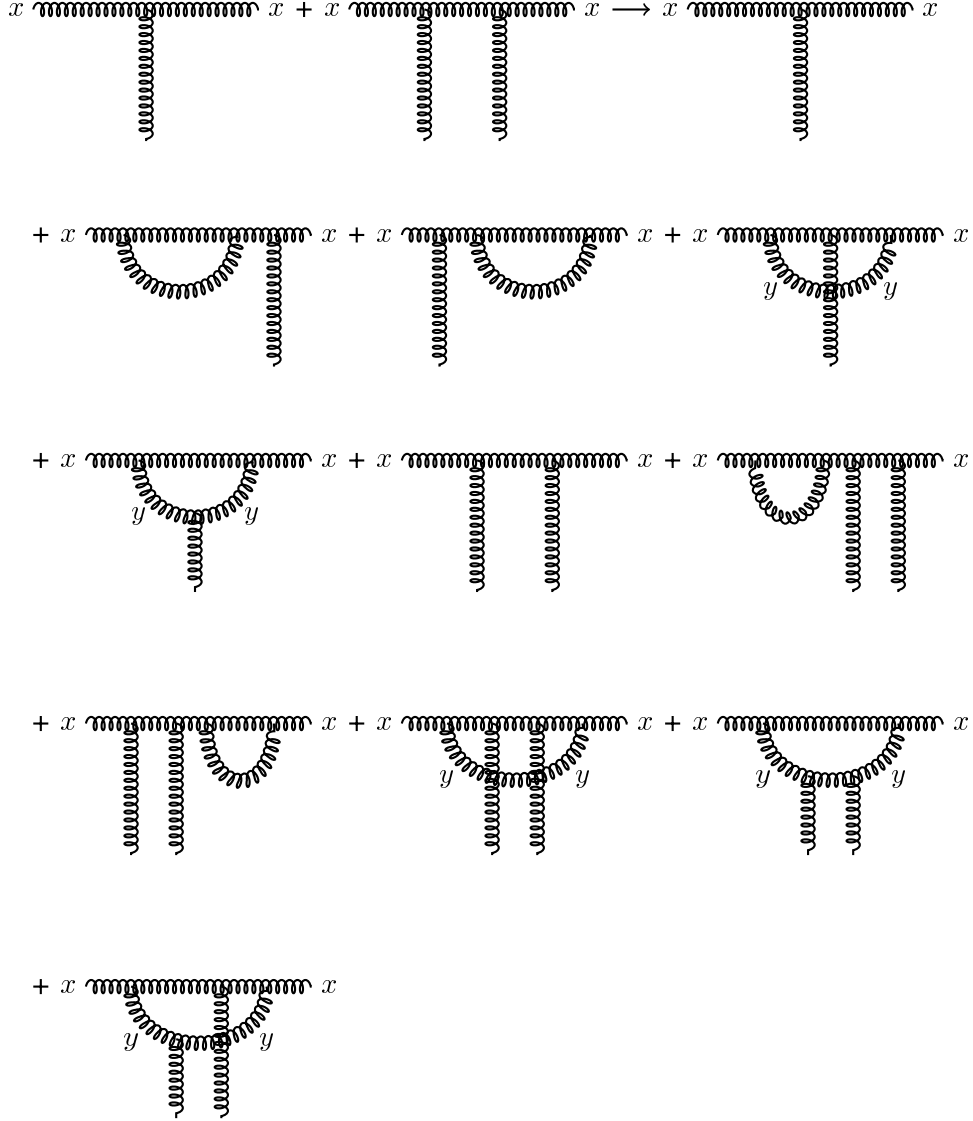}
\caption{\label{f2}KLWMIJ evolution in the approximation which allows exchanges of up to two gluons in the $t$-channel.}
\end{figure}

\begin{figure}
\includegraphics{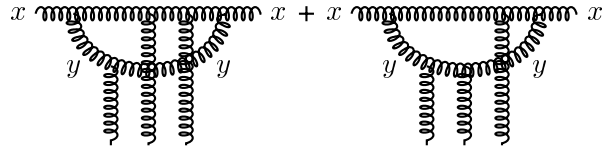}
\caption{\label{f3}A nonlocal in transverse plain contribution to the eigenfunction with three $t$ channel gluon exchanges.}
\end{figure}

As we have seen above, expansion in powers of $\delta/\delta\rho$ is the expansion in number of gluons exchanged in the $t$-channel. For example the linear approximation of eq.(\ref{regg}) allows only one gluon exchange between the partons of the projectile and the target. Diagrammatically the approximate diagonalization discussed in the previous section corresponds to summing the diagrams of Fig.\ref{f1}. The evolution therefore allows for emission of an arbitrary number of gluons in the wave function of the projectile, but only for a single gluon exchange between the evolved projectile and the target. The terms quadratic in $\delta/\delta\rho$ are represented in Fig.\ref{f2}. The vanishing of the last diagram on Fig.\ref{f2} leaves the remaining contributions local in transverse coordinate and thereby ensures the reggeization of the two gluon exchange. The third order terms are not local anymore. In particular one encounters the diagrams of Fig.\ref{f3} which give a non vanishing bilocal contribution.

%%%%%%%%%%%%%%%%%%%%%%%%%%%%%%%%%%%%%%%%%%%%%%%%%%%%%%%%%%%%%%%%%%%%
%%%%%%%%%%%%%%%%%%%%%%%%%%%%%%%%%%%%%%%%%%%%%%%%%%%%%%%%%%%%%%%%%%%%

%%%%%%%%%%%%%%%%%%%%%%%%%%%%%%%%%%%%%%%%%%%%%%%%%%%%%%%%%%%%%%%%%%%%
%%%%%%%%%%%%%%%%%%%%%%%%%%%%%%%%%%%%%%%%%%%%%%%%%%%%%%%%%%%%%%%%%%%%

%%%%%%%%%%%%%%%%%%%%%%%%%%%%%%%%%%%%%%%%%%%%%%%%%%%%%%%%%%%%%%%%%%%%
%%%%%%%%%%%%%%%%%%%%%%%%%%%%%%%%%%%%%%%%%%%%%%%%%%%%%%%%%%%%%%%%%%%%

%%%%%%%%%%%%%%%%%%%%%%%%%%%%%%%%%%%%%%%%%%%%%%%%%%%%%%%%%%%%%%%%%%%%
\begin{figure}
\includegraphics{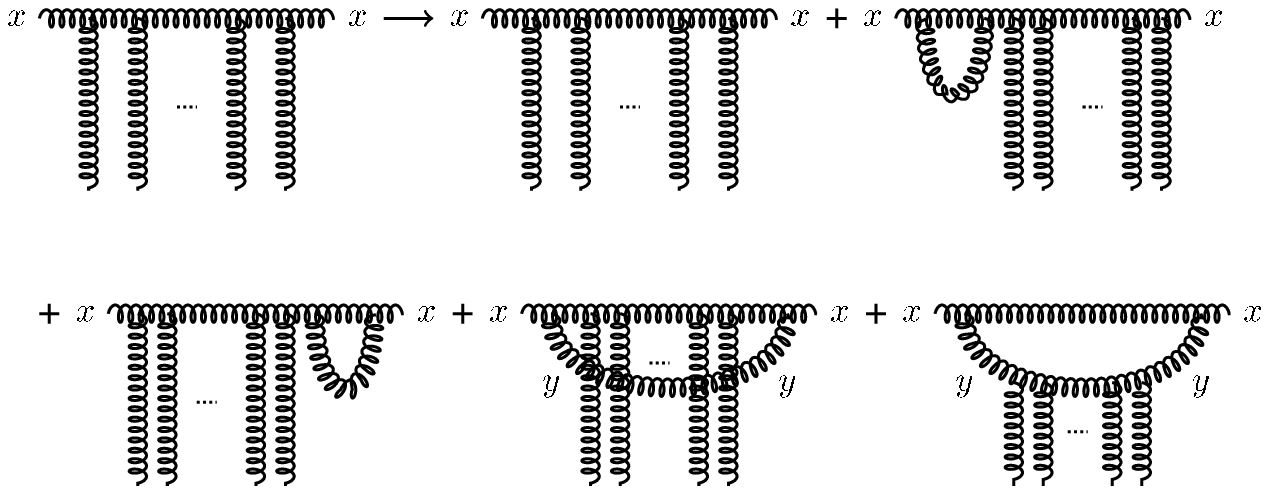}
\caption{\label{f4} Reggeization diagramms with arbitrary number of $t$-channel gluons that couple to the same parton in the projectile.}
\end{figure}
%%%%%%%%%%%%%%%%%%%%%%%%%%%%%%%%%%%%%%%%%%%%%%%%%%%%%%%%%%%%%%%%%%%%
Since the quadratic terms in the expansion of the wave function reggeize in the same way as the linear term, the significant corrections in a sense start form the cubic order. There is no reggeization of the third order terms in eq.(\ref{1glue}) nor eq.(\ref{regf}), which is another way of stating that three gluon exchange is sensitive to screening corections. It is thus interesting to calculate the terms of order $(\delta/\delta\rho)^3$ in the wave function. Although this can be done, the calculation is somewhat tedious. In this section we will perform a similar calculation, but the one that is easier implemented and has a somewhat more direct meaning in the framework of $H_{KLWMIJ}$. Instead of expanding the wave function $\Psi[R]$ in powers of $\delta/\delta\rho$ we will expand it in powers of $R-1$. This is similar in spirit to \cite{reggeon}. Since the complete KLWMIJ wave function must depend on $R$, this type of expansion is more direct.
To leading order the expansion in $R-1$ and $\delta/\delta\rho$ are equivalent. Beyond the leading order however they differ both on the calculational and conceptual levels. 

Expansion in powers of $R-1$ corresponds to the expansion in the number of the projectile partons which participate in scattering. To linear order in $R-1$ our approximation allows only one parton in the projectile wave function to scatter off the target, but it can scatter by exchanging an arbitrary number of $t$-channel gluons. The diagrams which are resummed in this approximation are depicted on Fig. 4.   Clearly the leading order expansion in $\delta/\delta\rho$ is subsumed in the leading order expansion in $R-1$, since if we allow only one $t$-channel gluon, we also allow only one parton of the projectile to scatter. In higher orders it is not the case anymore. In this section we will calculate $O[(R-1)^2]$ correction to the wave function of eq.(\ref{quark}).

Recall eq.(\ref{regf})
\begin{eqnarray}\label{hgaf}
&&H_{KLWMIJ}\int_u\phi_utr[\tau^aR_F(u)]=\frac{\alpha}{ 2\pi^2}\int_{u,z}K_{uuz}\phi_u\left\{tr[1-R^{\dagger}_{F}(z)R_{F}(u)]tr[\tau^aR_{F}(z)]+N_ctr\left[\tau^a\left(R_{F}(u)-R_{F}(z)\right)\right]\right\}\nonumber\\
&&=\frac{\alpha}{ 2\pi^2}\int_{u,z}K_{uuz}\phi_u\left\{N_ctr\left[\tau^a\left(R_{F}(u)-R_{F}(z)\right)\right]+tr[R_{F}(u)-1]tr[\tau^aR_{F}(z)]-tr[R^{\dagger}_{F}(z)-1]tr[\tau^aR_{F}(z)]+...
\right\}
\end{eqnarray}
The linear term on the RHS is the familiar reggeization term. We see thus that reggeization is the property of an arbitrary number of gluon exchanges, as long as all the $t$-channel gluons couple to {\bf the same parton} (in this case quark) in the projectile.
The second term on RHS is second order in $R-1$. Note that in terms of the expansion in $\delta/\delta\rho$ it actually starts with the cubic order. Clearly, to correct the eigenfunction we have to add to our original $G^a_F$ a term of second order in $R-1$. Guided by the form of the RHS of eq.(\ref{hgaf}) we take the wave function to second order of the form
\begin{eqnarray}
G^a_F&=&G^a_1+G^a_2\\
G^a_1&=&\int_u\phi_utr[\tau^aR_F(u)]\nonumber\\
G^a_2&=&\frac{1}{N_c}\int_{u,v}\psi(u,v)tr[R^{\dagger}_F(v)-1]tr[\tau^aR_F(u)]+\frac{1}{N_c}\int_{u,v}\tilde{\psi}(u,v)tr[R_F(v)-1]tr[\tau^aR_F(u)]\nonumber
\end{eqnarray}
We will see that this ansatz is general enough to satisfy the eigenvalue equation to second order in the large $N_c$ limit.

In the rest of this section all the matrixes $R$ are in fundamental representation and we drop the subscript $F$ for convenience.

The action of $H_{KLWMIJ}$ on $G$ is straightforward to calculate.  After some algebra we find

\begin{eqnarray}
&&H_{KLWMIJ}\int \psi(u,v)tr[R^\dagger_v-1]tr[\tau^aR_u]= \nonumber \\
&=&\frac{\alpha}{ 2\pi^2}\int_{u,v,z}\psi(u,v)\Bigg\{K_{uvz}\Big\{tr[R^\dagger_z R^\dagger_vR_z\tau^aR_u]+tr[R^\dagger_zR_u\tau^aR_zR^\dagger_v]-tr[R^\dagger_v\{R_u,\tau^a\}]\Big\}\nonumber \\&&
-K_{vvz}\Big\{tr(R^\dagger_z)tr[(R^\dagger_v-1)(R_z-1)]+tr(R^\dagger_z-1)tr(R^\dagger_v-1)\nonumber\\
&&+tr(R^\dagger_z-1)tr(R^\dagger_z-1)+N_c[tr(R^\dagger_z-1)+tr(R_z-1)]\Big\}tr(\tau^aR_u)\nonumber \\ &&-K_{uuz}\Big\{tr[R^\dagger_v-1]tr[(R^\dagger_z-1)(R_u-1)]+tr[R^\dagger_v-1]tr[R^\dagger_z-1]\nonumber\\
&&+tr[R^\dagger_v-1]tr[R_u-1]+N_ctr[R^\dagger_v-1]\Big\}tr[\tau^aR_u]\nonumber \\&&
+N_cK_{uuz}tr[R^\dagger_v-1]tr[\tau^aR_u]\Bigg\}
\end{eqnarray}
Keeping only $[O(R-1)^2]$ terms and taking large $N_c$ limit for simplicity, we have 
\begin{eqnarray}
&& H_{KLWMIJ}\int\psi(u,v)tr[R^\dagger_v-1]tr[\tau^aR_u]=  \\
&&=\frac{\alpha}{ 2\pi^2}\int_{uvz}N_c\Bigg\{\psi(u,z)K_{vvz}tr[1-R_v]tr[\tau^aR_u]+\Big[\psi(u,z)K_{vvz}+\psi(z,v)K_{uuz}-\psi(u,v)K_{uuz}\Big]tr[1-R^\dagger_v]tr[\tau^aR_u]\Bigg\}\nonumber
\end{eqnarray}

%\begin{eqnarray}
%&&\int K_{x,y,z}\tilde{\psi}(u,v)\left\{2J^c_L(x)[R_A(z)-1]^{cd}J^d_R(y)-[J^c_L(x)-J^c_R(x)][J^c_L(y)-J^c_R(y)]\right\}tr[R^\dagger_v-1]tr[T^aR_u]= \nonumber \\
%&=&\int_{u,v,z}\tilde{\psi}(u,v)\left\{K_{uvz}\left\{tr[R^\dagger_z R_uT^aR_zR_v]+tr[R^\dagger_zR_vR_zT^aR_u]-tr[R_v \{R_u,T^a \}]\right\}\right. \nonumber \\&&
%\left.+K_{vvz}\left\{tr(R_z)tr[(R^\dagger_z-1)(R_v-1)]+tr(R_z-1)tr(R^\dagger_z-1)+tr(R_z-1)tr(R_v-1)+N[tr(R_z-1)+tr(R^\dagger_z-1)]\right\}tr(T^aR_u)\right. \nonumber \\
%&&\left.+K_{uuz}\left\{tr[(R^\dagger_z-1)(R_u-1)]+tr[R^\dagger_z-1]+tr[R_u-1]+N\right\}tr[R_v-1]tr[T^aR_z]\right.\nonumber \\&&
%\left.-NK_{uuz}tr[R_v-1]tr[T^aR_u]\right\}
%\end{eqnarray}
Similarly to quadratic order in the large $N_c$ limit
%In the large $N_c$ limit this expression reduces to
\begin{eqnarray}
&&H_{KLWMIJ}\int tr[R_v-1]tr[\tau^aR_u]=  \\
&=&\frac{\alpha}{ 2\pi^2}\int_{uvz}N_c\left\{\tilde{\psi}(u,z)K_{vvz}tr[1-R^\dagger_v]tr[\tau^aR_u]+\left(\tilde{\psi}(u,z)K_{vvz}+\tilde{\psi}(z,v)K_{uuz}-\tilde{\psi}(u,v)K_{uuz}\right)tr[1-R_v]tr[\tau^aR_u]\right\}\nonumber
\end{eqnarray}
Finally we have
\begin{eqnarray}
HG^a_2&=&\frac{\alpha}{ 2\pi^2}\int_{u,v,z}\left\{\psi(u,z)K_{vvz}+\tilde{\psi}(u,z)K_{vvz}+\psi(z,v)K_{uuz}-\psi(u,v)K_{uuz}\right\}tr[1-R^\dagger_v]tr[\tau^aR_u]\nonumber\\
&+&\frac{\alpha}{ 2\pi^2}\int_{u,v,z}\left\{\tilde{\psi}(u,z)K_{vvz}+\psi(u,z)K_{vvz}+\tilde{\psi}(z,v)K_{uuz}-\tilde{\psi}(u,v)K_{uuz}\right\}tr[1-R_v]tr[\tau^aR_u]
\end{eqnarray}
We determine the functions $\psi$ and $\tilde\psi$ by solving the eigenvalue equation
\begin{equation}
H_{KLWMIJ}G^a_F=\omega_QG^a_F
\end{equation}
with the eigenvalue $\omega_Q$ given by eq.(\ref{intercept}). As we have seen in eq(\ref{hgaf}) the linear terms reggeize and cancel between the left and right hand side. For the quadratic terms we are then left with the equation

\begin{eqnarray}
&&\frac{\alpha}{ 2\pi^2}\int_{u,z}K_{uuz}\phi_u[tr(1-R^\dagger_z)+tr(1-R_u)]tr(\tau^aR_z)\nonumber \\
&+&\frac{\alpha}{ 2\pi^2}\int_{u,v,z}\left\{\psi(u,z)K_{vvz}+\tilde{\psi}(u,z)K_{vvz}+\psi(z,v)K_{uuz}-\psi(u,v)K_{uuz}\right\}tr[1-R^\dagger_v]tr[\tau^aR_u]\nonumber\\
&+&\frac{\alpha}{ 2\pi^2}\int_{u,v,z}\left\{\tilde{\psi}(u,z)K_{vvz}+\psi(u,z)K_{vvz}+\tilde{\psi}(z,v)K_{uuz}-\tilde{\psi}(u,v)K_{uuz}\right\}tr[1-R_v]tr[\tau^aR_u]\nonumber \\
&=&\frac{\omega_Q}{ N_c}\int_{u,v}[\psi(u,v)tr(R^\dagger_v-1)tr(\tau^aR_u)+\tilde{\psi}(u,v)tr(R_v-1)tr(\tau^aR_u)]
\end{eqnarray}
with $\phi_u=\exp \{iQu\}$.
This reduces to the equation for $\psi$ and $\tilde{\psi}$:
\begin{eqnarray}\label{correction}
\frac{\alpha}{ 2\pi^2}&&\int_z\left[\psi(u,v)K_{uuz}-\psi(u,z)K_{vvz}-\psi(z,v)K_{uuz}-\tilde{\psi}(u,z)K_{vvz}\right]-\frac{\omega_Q}{ N_c}\psi(u,v)=\frac{\alpha}{ 2\pi^2}\int_zK_{zzu}\phi_z\delta(u-v)\\
\frac{\alpha}{ 2\pi^2}&&\int_z\left[\tilde{\psi}(u,v)K_{uuz}-\tilde{\psi}(u,z)K_{vvz}-\tilde{\psi}(z,v)K_{uuz}-\psi(u,z)K_{vvz}\right]-\frac{\omega_Q}{ N_c}\tilde{\psi}(u,v)=\frac{\alpha}{ 2\pi^2}K_{vvu}\phi_v
\end{eqnarray}
These equations are easily solved in momentum space.
Defining
\begin{equation}
\psi(u,v)=\int d^2kd^2qe^{i(qu+kv)}\psi(q,k); \ \ \ \ \ 
\psi(q,k)=\int d^2ud^2ve^{-i(qu+kv)}\psi(u,v)
%\int d^2zK_{uuz}+w&=&\ln Q^2
\end{equation}
it is starightforward to Fourier transform eq.(\ref{correction}). For example
\begin{equation}
\int_{uv}e^{-i(qu+kv)}\int_zK_{vvz}\psi(u,z)= \ln \frac{\Lambda^2}{ k^2}\psi(q,k); \ \ \ \ \ etc.
\end{equation}
The ultraviolet cutoff $\Lambda$ is needed to regularize the divergence in the Fourier transfrom of $1/x^2$.
Similarly Fourier transforming the rest of the terms we get
\begin{eqnarray}
\left(\ln \frac{\Lambda^2}{ Q^2}-\ln \frac{\Lambda^2}{ q^2}-\ln \frac{\Lambda^2}{ k^2}\right)\psi(q,k)-\ln \frac{\Lambda^2}{ k^2}\tilde{\psi}(q,k)&=&\ln \frac{\Lambda^2}{ Q^2}\delta^2(Q-q-k) \\
\left(\ln \frac{\Lambda^2}{ Q^2}-\ln \frac{\Lambda^2}{ q^2}-\ln \frac{\Lambda^2}{ k^2}\right)\tilde{\psi}(q,k)-\ln \frac{\Lambda^2}{ k^2}\psi(q,k)&=&\ln \frac{\Lambda^2}{ q^2}\delta(Q-q-k)
\end{eqnarray}
%Defining $\psi(q,k)+\tilde{\psi}(q,k) \equiv \psi_+(q,k)$ and $\psi(q,k)-\tilde{\psi}(q,k) \equiv \psi_-(q,k)$
%\begin{eqnarray}
%\psi_+(q,k)&=&\frac{\ln{q^2Q^2}}{\ln{\frac{Q^2}{q^2k^4}}}\delta(Q-q-k) \\
%\psi_-(q,k)&=&\frac{\ln{\frac{Q^2}{q^2}}}{\ln{\frac{Q^2}{k^2}}}\delta^2(Q-q-k)
%\end{eqnarray}
%If one wants to find a symmetric and antisymmetric parts of $\Phi$ and $\tilde{\Phi}$ separetly , changing $q$ with $k$ in equations (70) and (71) and adding them up one can write 
%\begin{eqnarray}
%\tilde{\Phi}_a(q,k)=\frac{2\ln Q^2\delta(Q-q-k)}{\ln{\frac{q^2}{k^2}}} \\
%\Phi_a(q,k)= \frac{2\ln k^2q^2\delta(Q-q-k)}{\ln{\frac{k^2}{q^2}}} \\ 
%2\ln{\frac{Q^2}{q^2k^2}}\left(\Phi_s(q,k)+\tilde{\Phi}_s(q,k)\right)-\ln{q^2k^2}\left(\Phi_s(q,k)+\tilde{\Phi}_s(q,k)\right)&=&\left(2\ln Q^2+\ln{k^2q^2}\right)\delta(Q-q-k) \\
%2\ln{\frac{Q^2}{q^2k^2}}\left(\Phi_s(q,k)-\tilde{\Phi}_s(q,k)\right)-\ln{q^2k^2}\left(\Phi_s(q,k)-\tilde{\Phi}_s(q,k)\right)&=&\left(2\ln Q^2-\ln{k^2q^2}\right)\delta(Q-q-k) 
%\end{eqnarray}
The solution is found straightforwardly as
\begin{eqnarray}
\psi(q,k)&=&\frac{\ln \frac{\Lambda^2}{ Q^2}-\ln \frac{\Lambda^2}{ k^2}}{ \ln \frac{\Lambda^2}{ Q^2}-\ln \frac{\Lambda^2}{ q^2}-2\ln \frac{\Lambda^2}{ k^2}}\delta^2(Q-q-k)\rightarrow_{\Lambda\rightarrow\infty}0\nonumber\\
\tilde\psi(q,k)&=&\frac{\ln \frac{\Lambda^2}{ q^2}+\ln \frac{\Lambda^2}{ k^2}}{ \ln \frac{\Lambda^2}{ Q^2}-\ln \frac{\Lambda^2}{ q^2}-2\ln \frac{\Lambda^2}{ k^2}}\delta^2(Q-q-k)\rightarrow_{\Lambda\rightarrow\infty}-\delta^2(Q-q-k)
\end{eqnarray}
Transforming back to the configuration space we find the eigenfunction to second order in $R-1$ and in the large $N_c$ limit  
\begin{equation}
G^2_F=\int_ue^{iQu}tr[\tau^aR_F(u)]\Big[1-\frac{1}{ N_c}tr[R_F(u)-1]\Big]
\end{equation}

\section{Conclusions}
To summarize we have discussed relationship between the KLWMIJ Hamiltonian and its BFKL limit. Eigenfunctions of $H_{KLWMIJ}$ when expanded to leading order in $\delta/\delta\rho$ become eigenfunctions of $H_{BFKL}$. It is however difficult to determine which eigenfunctions of $H_{BFKL}$ become {\it normalizable} eigenfunctions of $H_{KLWMIJ}$ when the Taylor series is resummed to all orders.

The relation of course pertains only to functions (functionals) which are expandable in Taylor series in $\delta/\delta\rho$. We know from the general discussion of \cite{yinyang} that $H_{KLWMIJ}$ also has eigenfunctions which do not have such an expansion. Those are states close to the black disk limit.
For those states the pertinent expansion is in powers of $\rho$. We note that $H_{BFKL}$ is self dual under the transformation
\begin{equation}\label{duality}
\frac{\delta}{\delta\rho(x)}\leftrightarrow \int_y\frac{i}{\partial^2}(x-y)\rho(y)
\end{equation}
It thus contains eigenstates whose eigenfunctions are monomials in $\rho$ rather than $\delta/\delta\rho$. The self duality ensures that those have exactly the same spectrum as the eigenfunctions we discussed in the bulk of this paper, even though these states are indeed formally close to the black disk limit

The duality transformation eq.(\ref{duality}) is the linearized version of eq.(\ref{densedilute}) which transforms $H_{KLWMIJ}$ to $H_{JIMWLK}$. Thus the same relation as discussed above exists between the second set of the eigenfunctions of $H_{BFKL}$ and the eigenfunctions of $H_{JIMWLK}$.

We have also discussed the gluon reggeization and the appearance of the bootstrap condition in the KWLMIJ formalism. The bootstrap condition is direct consequence of Hermiticity of $H_{KWLMIJ}$ and as such is a necessary attribute of the approach. Any modification of the emission kernel $K_{xyz}$ does not ruin the bootstrap property.
Since the JIMWLK picture is a $s$-channel one while the reggeization stems from the  $t$ channel unitarity,
we conclude that the Hermiticity of $H_{KWLMIJ}$ is the property that reconciles these two approaches.

Further we have discussed expansion of the eigenfunctions in powers of $R-1$ rather than powers of $\delta/\delta\rho$. This corresponds to expansion in the number of partons in the projectile wave function which participate in the scattering. We have calculated $O[(R-1)^2]$ correction to the reggeized gluon wave function and found that is has a very simple form in the large $N_c$ limit.

We note in this respect that while the eigenvalue determines the rapidity dependence of the scattering amplitude, the functional form of the wave function determines the impact factor. This follows from the expansion eq.(\ref{psii}) which can be written as the overlap of the "wave function" characterizing the projectile hadron at the initial rapidity and the eigenfunction of the evolution Hamiltonian
\begin{equation}
\gamma_i=\langle W_{Y_0}|\Psi_s\rangle
\end{equation}
Thus for example the eigenfunction eq.(\ref{1glue}) will not have zero overlap with a single quark state, while that of eq.(\ref{quark}) will not overlap with a gluon projectile, since their incoming color representations cannot be combined into a singlet. Thus in order for the amplitudes of both, quark and gluon projectiles to have the same energy dependence, there must be a degeneracy in the spectrum of $H_{KLWMIJ}$. We indeed saw this degeneracy explicitly since both eigenfunctions eq.(\ref{1glue}) and eq.(\ref{quark}) correspond to the same eigenvalue. On the formal level this degeneracy is the consequence of the fact that perturbative vacuum breaks the global symmetry of $H_{KLWMIJ}$ as  $SU_L(N_c)\otimes SU_R(N_c)\rightarrow SU_V(N_c)$, the reggeized gluon being "the Goldstone boson" associated with this breaking \cite{reggeon}.

\section*{Acknowledgements} 

One of us (E.L.) thanks Physics Department of the University of Connecticut for hospitality and creative atmosphere during his visit when this work was done.

 This work was supported in part by the DOE grant DE-FG02-92ER40716.00 and  Fondecyt (Chile) grant  \# 1100648.

\section{Appendix. Derivation of $H_{BFKL}$ from $H_{KLWMIJ}$}
In order to derive the BFKL Hamiltonian we have to expand $R^{ab}(x)$ and $J^a_{L(R)}(x)$ in powers of  $\frac{\delta}{\delta \rho^a(x,x^-)}$. The expansion of $R$ is straightforward. To expand  $J_R$ and $J_L$ we will use their commutation relations with $R(x)$.  We start with the KLWMIJ Hamiltonian
\begin{equation}
H_{KLWMIJ}=\frac{\alpha}{ 2\pi^2}\int_zQ^{a \dagger}_i(z)Q^a_i(z)
\end{equation}
Expansion of $R$ is straightforward.  To second order we have
\begin{equation}
R^{\mu \beta}_u=\delta^{\mu \beta}+T^{\mu \beta}_c\int_0^1du^{-}_1\frac{\delta}{\delta \rho^c(u,u^{-}_1)}+T^{\mu \lambda}_cT^{\lambda \beta}_d\int_0^1du^{-}_1\int_0^{u^{-}_1}du^{-}_2\frac{\delta}{\delta \rho^c(u,u^{-}_1)}\frac{\delta}{\delta \rho^d(u,u^{-}_2)}
\end{equation}

To expand  $J_L$  we use the commutation relation 
\begin{equation}
[J^a_L(x),R(y)]=T^aR(y)\delta^2(x-y)
\end{equation}
It is easy to check that to second order in $\delta/\delta\rho$ the following expression satisfies the correct commutation relation
\begin{equation}
J^{a}_L (u) = -\int_0^1du^-_1\rho^{a} (u,u^{-}_1)-T^{\chi \kappa}_a\int_0^1du^{-}_1\int_0^{u^{-}_1}du^{-}_2\rho^{\chi}(u,u^{-}_2)\frac{\delta}{\delta \rho^{\kappa}(u,u^{-}_1)}
\end{equation}

Now using the fact that $J^a_L(u)=R^{ab}_uJ^b_R(u)$ we have also 
%\begin{equation}
%J^a_R(u)=-\int_0^1du^{-}_1\rho^a(u,u^{-}_1)-T^{\chi \kappa}_a\int_0^1du^{-}_1\int_0^{u^{-}_1}du^{-}_2\rho^\chi(u,u^{-}_2)\frac{\delta}{\delta \rho^\kappa(u,u^{-}_1)}+T^{\chi \kappa}_a\int_0^1du^{-}_1\int_0^1du^{-}_2\frac{\delta}{\delta \rho^\kappa(u,u^{-}_1)}\rho^\chi(u,u^{-}_2)
%\end{equation}
%Equvalently it can be written as
\begin{equation}
J^a_R(u)=-\int_0^1du^{-}_1\rho^a(u,u^{-}_1)+T^{\chi \kappa}_a\int_0^1du^{-}_1\int_{u^{-}_1}^1du^{-}_2\rho^\chi(u,u^{-}_2)\frac{\delta}{\delta \rho^\kappa(u,u^{-}_1)}
\end{equation}
Now the expansion of the amplitude $Q$ reads
\begin{eqnarray}
Q^a_i(z)&=&\int_x\frac{(x-z)_i}{(x-z)^2}\left[R^{ab}(z)-R^{ab}(x)\right]J^b_R(x) \nonumber \\
%&=&\int_x\frac{(x-z)_i}{(x-z)^2}\left[R^{ab}(z)J^b_R(x)-J^a_L(x)\right] \nonumber \\
&=&\int_x\frac{(x-z)_i}{(x-z)^2}T^{\alpha \beta}_a\left[\int_0^1dx^{-}_1\int_0^1dx^{-}_2\rho^\alpha(x,x^{-}_1)\frac{\delta}{\delta \rho^\beta(x,x^{-}_2)}-\int_0^1dx^{-}_1\int_0^1dz^{-}_1\rho^\alpha(x,x^{-}_1)\frac{\delta}{\delta \rho^\beta(z,z^{-}_1)}\right]\nonumber\\
&=&\int_x\frac{(x-z)_i}{(x-z)^2}T^{\alpha \beta}_a\left[\rho^\alpha_x\left(\frac{\delta}{\delta \rho^\beta_x}-\frac{\delta}{\delta \rho^\beta_z}\right)\right]
\end{eqnarray}
with
$\int_0^1dx^-\rho^\alpha(x,x^-)\equiv\rho^\alpha_x$ and $\int_0^1dx^{-}\frac{\delta}{\delta \rho^\alpha(x,x^-)}\equiv\frac{\delta}{\delta \rho^\alpha_x}$.
%\begin{equation}
%Q^a_i(z)=\int_x\frac{(x-z)_i}{(x-z)^2}T^{\alpha \beta}_a\left[\rho^\alpha_x\left(\frac{\delta}{\delta \rho^\beta_x}-\frac{\delta}{\delta \rho^\beta_z}\right)\right]
%\end{equation}
Finally we can write the BFKL Hamiltonian as in eq(\ref{bfkl}).
%\begin{equation}
%H_{BFKL}=K_{xyz}(T^aT^b)_{cd}\rho^a_x\left[\frac{\delta}{\delta \rho^c_x}-\frac{\delta}{\delta \rho^c_z}\right]\left[\frac{\delta}{\delta \rho^d_y}-\frac{\delta}{\delta \rho^d_z}\right]\rho^b_y
%\end{equation}

%
\end{document}